\documentclass[12pt]{article}
\usepackage{amsfonts}
\usepackage{amssymb,amsmath}
mm\evensidemargin=-5true mm \topmargin=-15true mm

\newcommand{\fr}[2]{\frac{#1}{#2}}

\newcommand{\gam}{\gamma}

\newcommand{\la}{\lambda}

\newcommand{\be}{\begin{equation}}
\newcommand{\ee}{\end{equation}}
\newcommand{\bea}{\begin{eqnarray}}
\newcommand{\eea}{\end{eqnarray}}

\def\ga{\mathrel{\raise.3ex\hbox{$>$\kern-.75em\lower1ex\hbox{$\sim$}}}}
\def\la{\mathrel{\raise.3ex\hbox{$<$\kern-.75em\lower1ex\hbox{$\sim$}}}}
\newcommand{\Mpl}{$M_{\rm {\small Pl}}$}

\newcommand{\ie}{{\it i.e.,}\ }
\newcommand{\eg}{{\it e.g.,}\ }
\newcommand{\reef}[1]{(\ref{#1})}
\newcommand{\cL}{{\mathcal L}}
\newcommand{\ssc}{\scriptscriptstyle}
\newcommand{\LL}[1]{\cL_{\ssc #1}}
\newcommand{\prt}{\partial}

\newcommand{\Psib}{\bar{\Psi}}
\newcommand{\veps}{\varepsilon}
\newcommand{\eps}{\epsilon}
\newcommand{\mpl}{M_{\rm {\small Pl}}}
\newcommand{\sla}[1]{\hbox{{$#1$}\llap{$/$}}}
\newcommand{\sixth}{{{\scriptstyle 1}\over{\scriptstyle 6}}}

\begin{document}

\setlength{\unitlength}{1mm}

\thispagestyle{empty} \rightline{\hfill \small gr-qc/0402028}
\vspace*{3cm}

\begin{center}
{\bf \Large Experimental Challenges
 for Quantum Gravity\footnote{Talk presented by {RCM}
at {\it QTS3: Third International Symposium on Quantum Theory and
Symmetries}, held at the University of Cincinnati, September
10-14, 2003. To appear in the Proceedings.}}\\
\vspace*{2cm}

{\bf ROBERT C.~MYERS}

\vspace*{0.2cm}

{\it Perimeter Institute for Theoretical Physics\\
35 King Street North, Waterloo, Ontario N2J 2W9, Canada\\
and\\
Department of Physics, University of Waterloo\\
Waterloo, Ontario N2L 3G1, Canada\\
E-mail: {\tt rmyers@perimeterinstitute.ca}}\\[.5em]

\vspace*{1cm}

{\bf MAXIM POSPELOV}

\vspace*{0.2cm}

{\it Department of Physics and Astronomy\\
University of Victoria, Victoria, BC, V8P 1A1, Canada\\
E-mail: {\tt pospelov@uvic.ca}}\\[.5em]

\vspace{2cm} {\bf ABSTRACT} 
\end{center}

The existence of a new fundamental scale may lead to modified
dispersion relations for particles at high energies. Such
modifications seem to be realized with the Planck scale in certain
descriptions of quantum gravity. We apply effective field theory
to this problem and identify dimension 5 operators that would lead
to cubic modifications of dispersion relations for Standard Model
particles. We also discuss other issues related to this approach
including various experimental bounds on the strength of these
interactions. Further we sketch a scenario where mixing of these
operators with dimensions 3 and 4 due to quantum effects is
minimal.

\vfill \setcounter{page}{0} \setcounter{footnote}{0}
\newpage


\section{Introduction}

The Planck mass, \Mpl, the dimensional parameter determining the
strength of gravitational interactions, remains a source of
conceptual problems for quantum field theories. When the momentum
transfer in two particle collisions is comparable to the Planck
mass, the graviton exchange becomes strong, signifying a breakdown
of the perturbative field theory description. Even without having
a fully consistent fundamental theory at hand, one can hypothesize
several broad categories of the low-energy effects induced by
\Mpl. The first group of such theories has only minor
modifications due to the existence of new physics at \Mpl. By
minor modifications we understand that all ``sacred'' symmetries
(Lorentz symmetry, CPT, spin-statistics, etc.) of the field theory
remain unbroken at low energies. Critical string theory in simple
backgrounds reduces to a field theory plus gravity at the scales
lower than $M_{s}$ and so provides an interesting example of this
category. In this case the chances to probe 1/\Mpl\ effects are
very remote as it is insufficient to merely have a probe with
large energy. Rather one must have extremely large momentum
transfers. Consequently, the propagation of a free particle with
large energy/momentum is immune to the effects of new physics, as
all corrections to the dispersion relation could be cast in the
form of $(p^2)(p^{2n}/M_{\rm Pl}^{2n})= (m^{2n+2}/M_{\rm
Pl}^{2n})$ where $m$ is the mass of the particle and $p$ is the
four-momentum.

The hope that nature could be more gracious to physicists is
reflected in the second class of scenarios, where the 1/\Mpl\
effects have much more ``vivid'' properties. Loop quantum gravity
seems to provide an example of this scenario. In this approach,
the discrete nature of space at short distances may be expected to
induce violations of Lorentz invariance and CPT. Such violations
are also often discussed in a broader context using field
theoretical language\cite{Kost}. Here one might assume a perfectly
Lorentz symmetric, CPT conserving action at $1/M_{\rm Pl}^{0}$
order and account for the existence of the Lorentz breaking terms
via a set of higher dimension operators. This would lead to
modifications of the dispersion relation for a free particle as
terms of the form $E^{n+2}/\mpl^{n}$ can appear. Such effects can
be searched for both with astrophysical
observations\cite{mass,rot,Jacob,wow,wow2} and with high precision
low-energy experiments\cite{terra,Vuc,prl}.

Cubic modifications of dispersion relations, which would appear at
the leading order in 1/\Mpl, have received considerable attention
in the literature
recently\cite{mass,rot,Jacob,wow,wow2,contrast,dsr}. In the
language of the effective field theory, such modifications can be
described by the dimension 5 operators. Although dimension 3 and 4
operators were extensively studied\cite{Kost}, dimension 5
operators remained unclassified and poorly explored. In our recent
work\cite{prl}, we took a first step towards such a classification
by considering dimension 5 operators in which Lorentz breaking can
be achieved by the introduction of a background four-vector $n^a$.
Within this framework, we studied cubic modifications of the
dispersion relation for scalars, vectors and fermions.
Specializing the operators to Standard Model particles, our
results showed that a cubic modification of the dispersion
relation is not possible for the Higgs particles, must have an
opposite sign for opposite chiralities of photons, and is
independent for different chiralities in the case of fermions.

This paper provides a summary of our earlier results\cite{prl}. We
begin with a review the class of dimension 5 operators which are
of interest to produce cubic modifications of dispersion
relations. We also discuss experimental constraints limiting the
strength of these operators. In particular, we demonstrate
remarkably stringent bounds\cite{prl} on these operators coming
from terrestrial clock comparison experiments.

Beyond this summary, we address further issues vis-a-vis Lorentz
violation and its effective field theory description in the
discussion section. In particular, we speculate on possible
mechanisms by which the violation of Lorentz invariance may be
restricted to higher dimensional operators and address the mixing
of the dimension 5 operators with those of dimension 3 and 4
through quantum effects. We are able to identify a scenario in
which this mixing may be minimal. We also argue that although our
previous discussion is framed in the language of Lorentz symmetry
breaking, many of the results still have application in scenarios
where the symmetry is deformed\cite{dsr}. Finally, we stress the
importance of a consistent formulation of Lorentz-violating (or
Lorentz-deforming) theories in the presence of
gravity\cite{gravity} and the necessity of correct identification
of the infrared degrees of freedom in such theories.

\section{Dimension 5 operators and cubic dispersion relations}

In the framework of low energy effective field theory, the
modified dispersion relations should be derived from some
appropriate modification of the kinetic terms in the Lagrangian.
Such modification may appear in the leading dimension 3 and 4
terms. However, we assume that short-distance physics does not
generate such Lorentz violating operators directly. From a
pragmatic point of view, this assumption is quite safe since even
if these terms exist, experimental constraints indicate that they
must be exceptionally small\cite{Kost}. Nevertheless, it poses
serious theoretical problems and we will attempt to address this
point in the closing discussion.

At the next level are dimension 5 operators which would lead to
$O(E^3)$ modifications of the dispersion relations. We adopt the
simplest approach where Lorentz symmetry is broken by a background
four-vector $n^a$ (with $n\cdot n=1$). We construct operators
satisfying six generic criteria:
\begin{enumerate}
\item Quadratic in the {\it same} field
\item One more derivative than the usual kinetic term
\item Gauge invariant
\item Lorentz invariant, except for the appearance of $n^a$
\item Not reducible to lower dimension by the equations of motion
\item Not reducible to a total derivative
\end{enumerate}
Conditions 2 and 5 ensure that these operators lead to $O(E^3)$
modifications of the dispersion relations, rather than $O(E^2m)$
or $O(Em^2)$, where $m$ is the mass of the particle. Our working
assumption will also be that these operators are naturally
suppressed by a factor of 1/\Mpl, and that $m\ll E \ll M_{\rm
Pl}$. This scaling ensures that all operators of dimension 5 can
be regarded as small perturbations. Below, we consider the cases
of vector and fermion particles. We refer the interested reader to
Ref.~\cite{prl} for the analogous discussion of scalars.

\noindent{\bf Vector:} Consider a U(1) gauge field with
the leading kinetic term, $\LL0 = - 
F^2/4$. The leading order equations of motion are just the
Maxwell equations, $\prt_a F^{ab}=0$. After gauge fixing
$\prt\cdot A=0$, this yields $\Box A_a=0$ or $k^2\,A_a(k)=0$ in
momentum space with $A_a\sim\exp(ik\cdot x)$. We wish to modify
the dispersion relation at $O(E^3)$ and so the new terms should
satisfy the constraints listed above. Keeping in mind the leading
order Maxwell equations and the Bianchi identities
$\prt_{[a}F_{bc]}=0$, one finds that there is a {\it unique} term
with the desired properties
\be \label{nextv} {\mathcal L}_\gamma= {\xi\over\mpl} n^a F_{ad}\,
n\cdot\prt(n_b\tilde F^{bd}), \ee
where $\tilde F^{ab}= \fr{1}{2} \veps^{abcd}F_{cd}$. Extension of this analysis to a
nonabelian vector is straightforward, and as in the abelian case
there is only one operator possible for each group at dim 5 level. Note that
operator (1) is odd under $CPT$ and even under charge
conjugation. The equation of motion becomes
\be\label{neweomv} \Box A_a={\xi\over\mpl}\veps_{abcd}\,n^b
(n\cdot\prt)^2F^{cd} \ee
again with the gauge choice $\prt\cdot A=0$. Further the right
hand side has been reduced through ample use of the Bianchi
identity. To identify the effect of the new term on the dispersion
relation, we go to momentum space and select photons moving along
the $z$ axis with $k^a=(E,0,0,p)$. Then for transverse
polarizations along the $x$ and $y$ axes,
\be \label{newdispv}
\left(E^2-p^2\pm{2\xi\over\mpl}p^3\right)(\eps_x\pm i\eps_y)
\simeq0 \ee
where we have used $E\simeq p$ to leading order and chosen the
``rest frame'' where $n^a=(1,0,0,0)$. Hence the sign of the cubic
term is determined by the chirality (or circular polarization) of
the photons. This leads to the rotation of the plane of
polarization for linearly polarized photons, which may be used to
bound $\xi$ \cite{rot,wow2}. Note that $\LL\gamma$ is unique and
hence the common approach\cite{mass,Jacob} of postulating a cubic
dispersion relation which is chirality independent is incompatible
with effective field theory.

\noindent{\bf Spinor:} Consider a Dirac spinor for which the
leading kinetic term is: $\LL0 = \Psib(i\sla{\prt}-m)\Psi.$ The
leading order equation of motion is just the Dirac equation:
$(\sla{\prt}+im)\Psi=0$. In momentum space with
$\Psi\sim\exp(-ik\cdot x)$, $(\sla{k}-m)\Psi(k)=0$. To modify the
dispersion relation at $O(E^3)$, we consider new terms satisfying
the constraints listed above. In this case, there are only two
terms with the desired form
\be \label{nextf} {\mathcal L}_f={i\over\mpl}\Psib\left(
\eta_1\sla{n}
+\eta_2\gam_5\sla{n}
\right)(n\cdot\prt)^2\Psi \ee
Both operators break CPT, with $\eta_1$ being charge conjugation
odd and $\eta_2$ charge conjugation even. After applying
$(i\sla{\prt}+m)$, the equation of motion takes the form
\be\label{neweomf2} (\Box+m^2)\Psi= {i\over\mpl}\left(\eta_1
+\eta_2\gam_5\right)(n\cdot\prt)^3\Psi\ .\ee
Here we have again dropped terms of order $m/\mpl$ or which vanish
by the leading order equations. Hence the modified dispersion
relation becomes
\be \label{newdispf}
\left(E^2-|\vec{p}|^2-m^2+{|\vec{p}|^3\over\mpl}(\eta_1
+\eta_2\gam_5)\right)\Psi =0 \ee
where we have used $E\simeq|\vec{p}|$ for high energies. At high
energies (\ie $E^2\gg m^2$), we can choose spinors as eigenspinors
of the chirality operator and redefine coupling constants as
$\eta_{L,R} = \eta_1 \mp \eta_2$. To introduce these operators for
Standard Model, the chiral choice for $\eta$ couplings would be
required by gauge invariance. Previous studies\cite{Jacob,wow}
considered only chirality independent dispersion relations for
fermions and so implicitly fix $\eta_2=0$.

Above we have identified interesting operators which modify the
dispersion relations at cubic order for vectors or fermions. The
external tensor appearing in all of these operators takes the form
$n^an^bn^c$. For technical reasons to be addressed in the
discussion section, we must replace this coupling by the traceless
symmetric tensor $C^{abc}= n^an^bn^c -\sixth$($n^ag^{bc}+$cyclic)
in the following section. Note that this change implicitly
introduces extra terms which do not satisfy all of the constraints
listed above, \ie they are reducible by the leading order
equations of motion. However, this replacement does not affect the
dispersion relations in the regime $E\gg m$. One could also
consider frame-dependent modifications of the interaction terms
between, \eg photons and electrons. If we limit ourselves to the
traceless symmetric coupling, $C^{abc}$, introduced here, there
are in fact no additional interaction terms with dimension 5
beyond those implied by extending Eq.~(\ref{nextf}) with
gauge-covariant derivatives.

\section{Experimental constraints on dimension 5 operators}

Evidence of modified dispersion relations for stable particles
such as electrons, light quarks, and photons can be searched for
using the astrophysical probes\cite{mass,rot,Jacob,wow,wow2}. We
begin here, however, by showing that impressive constraints can be
imposed by considering terrestrial experiments. These indirect
limits exploit the idea that the external four-vector $n^a$
introduces a preferred frame that can not coincide with the
laboratory frame on the Earth\cite{nuCPT}. While the component
$n^0$ is still dominant, the motion of the galaxy, Solar system
and Earth will create spatial components $n_i \sim 10^{-3}$ for a
terrestrial observer. Hence clock comparison
experiments\cite{clocks} or searches for spatial
anisotropy\cite{Heckel} can impose stringent bounds on violations
of Lorentz symmetry in this context. This approach was recently
used to constrain the dispersion relation for nucleons\cite{Vuc}.
Our constraints\cite{prl} apply to the fundamental fields of the
Standard Model rather that presumably should have more direct
connection to the Planck scale physics than nucleons.

Limits on the operators involving electrons and electron neutrinos
are especially easy to derive. We use the fact that best
tests\cite{Heckel} of directional sensitivity in the precession of
electrons limit the size of interaction between the external
direction and the electron spin at the level of $10^{-28}$ GeV.
This immediately translates into the following limit on the
coefficients $\eta_L$ and $\eta_R$ that parametrize the effective
interaction of the form \reef{nextf} for left-handed leptons and
right-handed electrons\cite{prl}:
\be |\eta_L^e-\eta_R^e|\la \fr{10^{-28}~{\rm
GeV}\mpl}{m_e^2|n_i|}\simeq 4, \label{leptl} \ee
where $\mpl \equiv 10^{19}$ GeV. The combination,
$\eta_L^e+\eta_R^e$, would be very weakly constrained here, as it
does not appear in the electron spin Hamiltonian.

The absence of a preferred direction is checked with even greater
precision using nuclear spin, which translates into more stringent
limits on new operators for the light quarks. The photon operator
(\ref{nextv}) will also contribute because of the electromagnetic
interactions inside the nucleon. To use the best experimental
limits of $10^{-31}$ GeV on the coupling of $n_i$ to neutron
spins\cite{clocks}, we must relate the photon and quark operators
with nucleon spin. First, let us introduce dimension 5 operators
for the first generation of quarks, the left-handed doublet
$\psi_Q$ and right-handed singlets $\psi_u$ and $\psi_d$:
\bea {\mathcal L}_q &=& \frac{C^{abc}}{\mpl}
\sum_{i=Q,u,d}\eta_i\, \bar \psi_i \gamma_a \prt_b\prt_c \psi_i\ .
\label{Qud} \eea
At the nucleon level, we estimate using standard QCD sum rules
\begin{eqnarray}
\label{matel}
\eta_{1,N} = a_{u}(\eta_u +\eta_Q)+a_{d}(\eta_d +\eta_Q)\\
\eta_{2,N} =b_{u}(\eta_u -\eta_Q)+b_{d}(\eta_d -\eta_Q) +
b_{\gamma}\xi, \nonumber \eea
where $\eta_{1(2),N}$ are the $\eta_1$ and $\eta_2$ couplings for
nucleons defined in (\ref{nextf}). Note that $\xi$ enters only in
the $\eta_2$ coupling for nucleons because both are even under the
charge conjugation. In (\ref{matel}), $a_{u,d}$ and $b_{u,d}$ are
the matrix elements that could be obtained as the moments of the
experimentally measured structure functions\cite{structure}: $a_d
\sim 0.4 $, $a_u \sim 0.1 $, $b_d\sim 0.1 $, $b_u\sim -0.05$ for
the neutron and charge inverted values for the proton. To relate
the photon operator with nucleon, we use the simplest vector
dominance model and obtain at one-loop level $b_\gamma \sim
0.13\alpha/(4 \pi)$ for neutron and $b_\gamma \sim 0.24 \alpha/(4
\pi)$ for proton. Combining these results produces the following
limit:
\begin{eqnarray}
|(\eta_d - \eta_Q) - 0.5 (\eta_u - \eta_Q) + 10^{-3}\xi| \la
10^{-8} \label{etaquark}
\end{eqnarray}
Barring accidental cancellations, we can place separate limits on
$\eta_{u,d} - \eta_Q$ at $10^{-8}$ and on $\xi$ at $10^{-5}$
level. The orthogonal combinations $\eta_{u,d} + \eta_Q$ are less
constrained because they enter only in the quadrupole coupling
between the nuclear spin and external direction\cite{Vuc}, and
thus are suppressed by an additional factor of $|n_i|\sim
10^{-3}$.

So far we have neglected the fact that the low energy values for
the couplings $\eta_i$ and $\xi$ taken at the normalization scale
of 1 GeV do not coincide with the high-energy values for the same
couplings generated at $\mpl$. With a simple one-loop analysis of
the renormalization group equations, we find that several bounds
can be strengthened. Leaving the details for elsewhere\cite{more},
our results are: $|\eta_{Q,u,d}|,|\xi| \la 10^{-6}$ and
$|\eta_{L,R}^e| \la 10^{-5}$.
The constraints may also be improved by assuming degeneracies
appropriate for grand unification at the GUT scale. On the
experimental side, the recent progress\cite{Romalis} in the
high-sensitivity atomic magnetometers that are unaffected by
spin-exchange relaxation may lead to an improvement of terrestrial
bounds on Lorentz violation by several orders of magnitude.

In summary, we have shown that effective field theory provides a
framework where one can derive stringent bounds on Planck scale
interactions from terrestrial experiments. The resulting limits
enhance and generalize the terrestrial bounds obtained
previously\cite{Vuc}. They are generally far more sensitive than
those previously derived by considering astrophysical
phenomena\cite{Jacob} where the typical sensitivity is $O(1)$ in
units of $\mpl^{-1}$. A notable exception is the bound\cite{rot}
$|\xi|\la2\times10^{-4}$ derived from the birefringence induced by
$\LL\gamma$. Our bound\cite{prl} improved on this result by
roughly an order of magnitude. However, with new astronomical
data, this bound was recently updated\cite{wow2} in a striking way
to $|\xi|\la10^{-14}$.

Our limit \reef{leptl} is already comparable to previous
constraints from the astrophysical searches of vacuum photon decay
and the absence of the vacuum Cerenkov radiation\cite{Jacob}.
Further, the latter analysis did not consider polarization effects
(\ie assumed $\eta_2=0$) and so this result provides a
complementary constraint. Certainly our bounds inferred from the
renormalization group analysis\cite{prl,more} represent a major
improvement in constraining these couplings. However, these bounds
were already tightened to an extraordinary level by other recent
work\cite{wow}. There, the observation of synchrotron radiation
from the Crab nebula was used to infer
$\eta_{L}^e+\eta_{R}^e\ga-10^{-7}$. We should also note that
combining various astrophysical constraints may yield $O(10^{-2})$
constraints on the difference $\eta_R^e-\eta_L^e$ subject to
certain assumptions\cite{wow2}.

\section{Discussion}

The procedure of classifying operators according to their
dimension proves very useful in analyzing sensitivity of various
experiments to Lorentz violation. In such an analysis, one should
always start from the {\it lowest} possible operators that may be
induced by the short-distance physics. In the case of Lorentz
violation, this expansion starts from dimension 3 \cite{Kost}, and
hence we arrive at a certain ``naturalness" problem for the whole
approach: Why should the size of dimension 3 operators not be
$\mpl$, the scale of UV physics responsible for the breaking,
rather than less than $\sim 10^{-30}$ GeV as various experimental
tests of Lorentz invariance demand? Similarly, the naive
expectation would be that Lorentz violation should appear with
$O(1)$ strength at dimension 4, again in contradiction with
experimental constraints. Perhaps, the full answer to this puzzle
requires a proper understanding of the full dynamical picture of
Lorentz violation emerging from high-energy scales. Without that
we have to resort to a number of plausible explanations, none of
them very compelling at present.

One possbility is that the properties of the theory at the UV
scale are inconsistent with the presence of dimension 3 and 4
Lorentz-violating operators. For example, it can be easily argued
that the exact supersymmetry at UV scale may completely forbid or
at the very least severely restrict the possibility of dimension 3
and 4 operators. A classification of all supersymmetric Lorentz
non-invariant operators is an interesting study on its own, but
goes outside of the present discussion\cite{susy}.

Another generic possibility to entertain in this approach is that
the properties of the Lorentz-violating background itself lead to
the suppression/exclusion of dimension 3 and 4 operators. For
example, one might consider Lorentz violations as realized by a
symmetric traceless rank-three tensor, $C^{abc}$, as was
introduced at the end of section 2. In euclidean space, such
tensor would correspond to an intrinsic {\it octupole} deformation
of the vacuum. One finds that there are simply no dimension 3 or 4
operators which couple to this tensor. Beyond an absence of bare
couplings, this implies that in a leading order calculation linear
in $C^{abc}$ can not generate dimension 3 or 4 operators at any
number of loops\cite{more}. This was of particular importance in
section 3 where we found the dimension 5 operators evolved
logarithmically and while dimension 3 operators might have been
expected to appear with $\Lambda_{UV}^2/\mpl$ coefficients, they
were in fact absent. Setting $C^{abc}= n^an^bn^c
-\sixth$($n^ag^{bc}+$cyclic) was a simplifying ansatz in our
analysis and is not required by any underlying principles. Notice
that if the dimension 5 operators coupled to $n^an^bn^c$ rather
than $C^{abc}$, one-loop graphs already generate, \eg
$\Lambda^2_{UV}/\mpl\ \bar \psi \sla{n}\gamma_5\psi$. The
appearance of such quadratic divergence results because the new
coupling is not in an irreducible representation of the Lorentz
group and, as well as $C^{abc}$, contains a vector component which
can couple to dimension 3 and 4 operators.

One can also consider higher order quantum corrections beyond
linear order in $C^{abc}/\mpl$. A priori, there is no reason to
believe that this effect should be small. For example, a triple
product $C_{ab}^{c}C^{abd}C_{cd}^e$ will transform as a vector,
and therefore may generate dimension 3 operators with a
coefficient $\Lambda_{UV}^4/\mpl^{-3}$. If the scale of the
ultraviolet cutoff in the loops is itself of the order of the
Planck scale, then the resulting size of dimension 3 operator
would be too large\cite{more,comment}. Therefore, one is forced to
assume that an effective cutoff in the loops with $\Lambda_{UV}
\ll \mpl$. If we require these induced dimension 3 operators to be
smaller than dimension 5 operators at energy scales of 1 GeV, we
must assume that $ \Lambda_{UV} \la \sqrt{\mpl\times 1{\rm GeV}}
\sim 10^9$GeV. This scale is far larger than the electroweak scale
and/or the scale of the supersymmetry breaking. Even though a
generic investigation of the supersymmetric stabilization of the
ultraviolet divergencies in the presence of the Lorentz violating
operators is presently lacking, it certainly appears as a
plausible scenario\cite{susy}. To summarize, one must regard both
the appearance of $C^{abc}$ and a low scale cutoff in the loops as
essential ingredients in suppressing the appearance of Lorentz
violations in lower dimension operators from quantum effects.

Following Ref.~\cite{Vuc}, it is amusing to compare the
experimental constraints to semi-classical calculations which have
appeared in the loop quantum gravity
literature\cite{loop1,loop2,loop3}. For a Dirac fermion, these
studies\cite{loop2} suggest that $\eta_1=0$ while $\eta_2$ is
nonvanishing. However, the latter is suppressed by factors of
M$_{\rm L}/\mpl$ where M$_{\rm L}\ll\mpl$ is the coherence scale
of the gravitational wave function. Therefore these calculations
seem to be in agreement with the stringent experimental
bounds\cite{wow,wow2,prl} imposed on the fermion operators. In
contrast, a separate analysis\cite{loop1} suggests that $\xi$
should be $O(1)$, which stands in stark contradiction with the
bounds discussed above\cite{rot,wow2,prl}. Hence naively the
experimental bounds seem to be in conflict with the predictions of
loop quantum gravity. However, it must be noted that the latter
results are at present somewhat heuristic and so this apparent
contradiction should not be taken too seriously.

Further recent calculations\cite{recent} suggest that the effects
of quantum gravity may deform the Lorentz symmetry, \ie departures
from the ``standard" realization of Lorentz symmetries may be
induced by Planck scale effects. This scenario, commonly known as
``doubly special relativity" (DSR)\cite{dsr}, would seem a drastic
departure from Lorentz breaking, in which our discussion was
phrased. However, we will argue that many of the results are still
applicable in this new scenario.

For simplicity, focus on photons and electrons which are central
to many of the discussions of experimental bounds. First one must
assume that the present preliminary investigations of DSR can be
extended to provide something like a quantum field theory of
photons and electrons incorporating DSR. Now the first simple
observation is that in the limit where \Mpl$\rightarrow\infty$,
this DSR theory must reduce to ordinary QED. Then for finite \Mpl,
the low-energy or long-wavelength physics can still be described
using standard techniques of effective field theory. Hence if the
dispersion relations relevant for photons and electrons are
modified at $O(p^3)$ (and we assume that rotational invariance is
unmodified), then it must be by the appearance of dimension 5
interactions of the form given in Eqs.~\reef{nextv} and
\reef{nextf}. In this context, one should not think that these new
interactions break Lorentz invariance. Rather the apparent Lorentz
violations induced by standard Lorentz transformations on these
terms must be compensated by corrections to these transformations
when they act on the standard kinetic terms. In fact, the latter
observation may be used as a strategy to infer the leading
corrections to the Lorentz transformations\cite{unpub}. In any
event, the discussion of section 2 is equally applicable to DSR
scenarios as to scenarios of Lorentz breaking.

One must, of course, be more cautious when considering the
experimental bounds quoted above, as many of these rely on the
existence of a preferred frame\cite{contrast}. However, we argued
above that Eq.~\reef{nextv} still gives the leading modification
of the photon action in a DSR scenario, and hence the $O(p^3)$
modification of the dispersion relation is photon-chirality
dependent.\footnote{It is curious that certain arguments seem to
imply this would be inconsistent with DSR\cite{contrast}.}
Further, the astronomical tests involving the resulting
birefringence for photons provide purely kinematical constraints
on the theory. That is, one relies on the magnification of small
but unusual effects in the propagation of photons as they cross
cosmological distances, without reference to a preferred frame.
Hence these bounds\cite{rot,wow2} are again applicable to DSR
scenarios. Given the present experimental bound\cite{wow2} that
$|\xi|\la10^{-14}$, it seems that cubic modifications of the
photon dispersion relation are essentially ruled out for DSR
constructions.

Another important aspect of the effective field theory approach to
Lorentz violation is the correct identification of the infrared
degrees of freedom and the possibility of generalizing this
framework to general relativity\cite{gravity}. Attempts to
generalize Lorentz breaking operators to general relativity would
necessarily supply $C^{abc}$ with coordinate dependence and a
corresponding kinetic term, which in turn would contribute to the
energy density. There are several arguments supporting the idea
that Lorentz violating tensors must be accompanied by massless (or
nearly massless) particles in the spectrum. Many explicit examples
involving Lorentz violation point towards this result, including
noncommutative backgrounds in string theory\cite{nonc},
super-light axion backgrounds\cite{carroll} and ghost
condensation\cite{nima}. The exchange by new light degrees of
freedom may lead to the appearance of a new coherent interaction,
\ie a ``fifth force". Therefore, quite separate bounds on Lorentz
violation may emerge from gravitational physics and cosmology.
Elucidating this question is also important for interpreting the
bounds on Lorentz violation that come from cosmological distances,
over which a ``constant" background would most certainly be
changing.

In conclusion, we would like to stress again that there remains
much work to be done both on the theoretical description from
quantum gravity and the phenomenological constraints. However, it
is truly remarkable that present-day precision experiments and
astrophysical observations can already confront quantum gravity
calculations with concrete and stringent observational bounds.



\section*{Acknowledgments}

RCM would like to thank the organizers of QTS3 for the opportunity
to present this material. We would like to thank Giovanni
Amelino-Camelia, Cliff Burgess, Laurent Freidel, Stefan Groot
Nibbelink, Ted Jacobson, Jerzy Kowalski-Glikman, Joe Lykken, Seth
Major, Guy Moore, Ann Nelson, Adam Ritz, Mike Romalis, Subir
Sarkar and Lee Smolin for many useful comments and conversations
over the course of this work. This research is supported in part
by NSERC of Canada, Fonds FCAR du Qu\'ebec and PPARC UK.







\end{document}